\documentclass[prl,twocolumn,showpacs]{revtex4}
\usepackage{amsfonts}
\usepackage{amsmath}
\usepackage{amssymb}
\usepackage{graphicx}
\usepackage{graphics}

\begin{document}

\author{V. Berejnov}
\address{Physics Department, Cornell University, Ithaca, NY, 14853-2501}
\author{A.Leshansky}
\address{Chemical Engineering Department,Technion, Haifa, Israel, 32000}

\noindent\textbf{Comment on "Self-Running Droplet: Emergence of Regular Motion from
Nonequilibrium Noise"} \\

In a recent Letter \cite{Sumino} a spontaneous motion of an oil
droplet in the surrounding aqueous media along a glass surface was
reported. The authors suggested that the \emph{self-locomotion} is
driven by a difference in wettability ($\Delta\gamma_{sl}$)
between the front and the rear of a droplet sustained by its
translational motion. Whereas we do not question the presented
experimental evidences, our comment concerns their interpretation
and the underlying physical mechanism of the motion.

We have analyzed movies 2 and 3 (Ref.22 in \cite{Sumino}) in
details using LabView software. For every frame of the
de-fragmented movies both advanced and receding contact regions of
the drop's contour ($\sim 50$ points) were subtracted and
approximated by polynomial functions. The drop contour was then
extrapolated to the substrate and the contact angles and
instantaneous velocities were determined.

\begin{figure}[h]
\qquad
\includegraphics[width=8.4 cm]{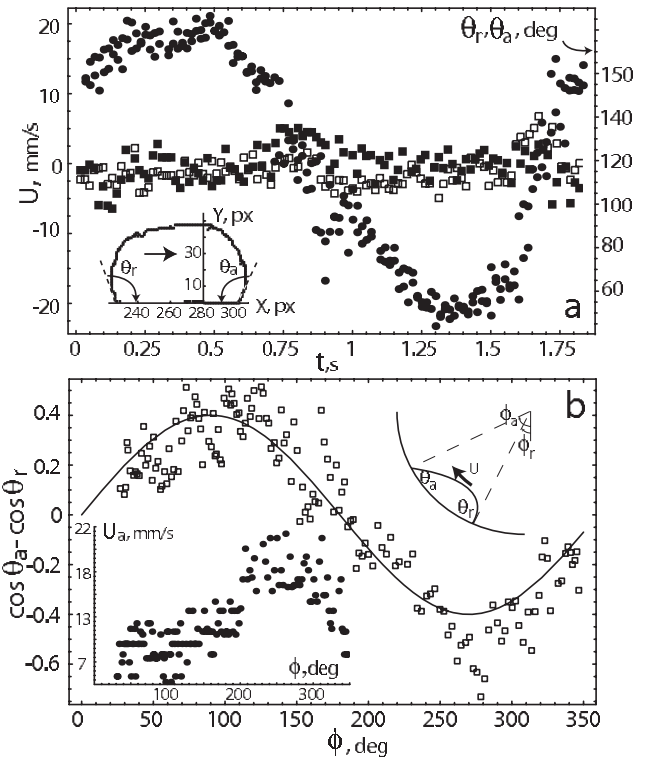}
\caption{(a) Horizontally moving drop (movie 2), $\theta_a$
($\blacksquare$), $\theta_r$ ($\square$) and $U$ ($\bullet$); (b)
The drop moving inside the vertical circle (movie 3),
$\cos\theta_a-\cos\theta_r$ ($\square$), $U$ ($\bullet$) and
prediction of \cite{Frenkel} (\hbox{---$\!$---}).}\label{fig1}
\end{figure}

For repetitive motion of  a droplet on a narrow horizontal strip
(movie 2, Fig.\ref{fig1}a), the mean values of the advancing and
receding contact angles are $\theta_a=(115.7\pm5.4)^\circ$,
$\theta_r=(113.5\pm3.5)^\circ$, respectively, and
$\Delta\theta=(2.2\pm6.4)^\circ\sim0$ that substantially deviates
from $11^\circ$ presented in \cite{Sumino}. The instantaneous
velocity $U(t)$, exhibits sinusoidal behavior without any
appreciable variation in $\Delta\theta$ in time, that is also
inconsistent with \cite{Sumino}.

The analysis of a droplet motion inside the vertical circle (movie
$3$, Fig.\ref{fig1}b) shows a variation of $\Delta\theta$ vs. the
angular position $\phi$ similar to that caused by a
\textit{static} interplay between gravity and capillarity in
presence of the contact angle hysteresis \cite{Frenkel}. If the
motion was driven by $\Delta\gamma_{sl}$ between the front and the
rear of the drop, one would expect $\cos\theta_a-\cos\theta_r$ to
deviate noticeably from the prediction of \cite{Frenkel} (the
solid line in Fig.\ref{fig1}b).

We have successfully reproduced most of the experiments of
\cite{Sumino} with a similar system: Wt/CTAB(0.45--1.0 mM) and benzonitrile/NaI(8.0--80 mM).
We also performed a similar experiment with oil lenses
deposited at the air-water interface \footnote{Replacing nitrobenzene
(SG$\sim1.2$) used in \cite{Sumino} with
benzonitrile (SG$\sim1.01$) allows deposition of lenses of 160 $\mu$l
at the a/w interface.}. The observed motion and the speed (up to
$\sim$$1$ cm/s) of the self-running lenses were similar to those
measured for immersed drops in \cite{Sumino}. Thus, interaction
with solid surface is not central for self-locomotion.

We argue that the mechanism of propulsion is that due to
\emph{chemo-capillary} effect. It has been predicted in
\cite{Golovin} that Marangoni convection may result in a
spontaneous motion of a reactive droplet suspended in a viscous
fluid. Experimentally observed velocities of chemo-capillary drift
cover a wide range from $~0.1$ mm/s \cite{Kosv} to $~1$ cm/s
\cite{ChifuBekki}. We found that drop interface consists of
domains with low surface tension, $\gamma$. The transport across
the w/o interface in the aria of those domains is accompanied by
the formation of \emph{microemulsion} phase inside the drop.
\cite{lowgam}.
It is well known that $\gamma$, between the microemulsion phase
and the co-existing water-rich phase is ultralow, \textit{i.e.},
$3$-$4$ orders of magnitude smaller than the bare oil-water value.
Thus, the gradients of $\gamma$ in the vicinity of the domain's
fronts could be enormously large \cite{ChifuBekki}, yielding
$\sim$cm/s speeds, while in \cite{Kosv} the motion is driven by
weak near-equilibrium variation of $\gamma$ with surface
concentration of surfactant. \\

\noindent V. Berejnov$^1$ and A. Leshansky$^2$\\
$^1$Physics Department, Cornell University, Ithaca, NY,
14853-2501; vb54@cornell.edu\\ $^2$Chemical Engineering
Department, Technion, Haifa, Israel, 32000;
lisha@techunix.technion.ac.il \\


\begin{thebibliography}{99}
\bibitem{Sumino} Y. Sumino \textit{et al.,} Phys. Rev. Lett. \textbf{94}, 068301 (2005).
\bibitem{Frenkel} Y. I. Frenkel, J. Exptl. Theoret. Phys. \textbf{18}, 659,
(1948); \\
(english translation: xxx.lanl.gov/abs/physics/0503051).
\bibitem{Golovin} A. A. Golovin and Yu. S. Ryazantsev, Fluid Dynamics \textbf{23},
370 (1990).
\bibitem{Kosv} S. R. Kosvintsev \textit{et al.,} Colloid Journal \textbf{63}, 318 (2001).
\bibitem{ChifuBekki} E. Chifu \textit{et al.,} J. Colloid Interface
Sci. \textbf{93}, 140 (1983); S. Bekki \textit{et al.,} J. Colloid
Interface Sci. \textbf{152}, 314 (1992);
\bibitem{lowgam} A. M. Bellocq \textit{et al.,} J. Colloid Interface Sci. \textbf{89}, 427 (1982);
A. M. Cazabat \textit{et al.,} Adv. Colloid Interface Sci.
\textbf{16}, 175 (1982).

\end{thebibliography}
\end{document}